\documentclass[journal]{IEEEtran}
\usepackage{xcolor}
\usepackage{bm}

\usepackage{cite}
  \usepackage{graphicx}
\usepackage[cmex10]{amsmath}
\usepackage{algorithmic}
\usepackage{array}
\usepackage[caption=false,font=normalsize,labelfont=sf,textfont=sf]{subfig}
\usepackage{fixltx2e}
\usepackage{stfloats}
\begin{document}

\title{Advances in 3D scattering tomography of cloud micro-physics}

\author{Masada~Tzabari,
        Vadim~Holodovsky,
        Omer~Shubi,
        Eitan~Eshkol,
        and~Yoav~Y.~Schechner% <-this % stops a space
\thanks{M. Tzabari, V. Holodovsky, O. Shubi, and Y. Y.  Schechner are with the Viterbi Faculty of Electrical Engineering, Technion - Israel Institute of Technology, Haifa, Israel e-mail: masada.tz@campus.technion.ac.il.}% <-this % stops a space
\thanks{E. Eshkol is with the Department of Earth and Planetary Sciences, The Weizmann Institute of Science, Rehovot, Israel.}% <-this % stops a space
\thanks{M. Tzabari and V. Holodovsky contributed equally to this work as first authors.}}

\maketitle

\begin{abstract}
We introduce new adjustments and advances in space-borne 3D volumetric scattering-tomography of cloud micro-physics. The micro-physical properties retrieved are the liquid water content and effective radius within a cloud. 
New adjustments include an advanced perspective polarization imager model, and the assumption of 3D variation of the effective radius. 
Under these assumptions, we advanced the retrieval to yield results that (compared to the simulated ground-truth) have smaller errors than the prior art. Elements of our advancement include initialization by a parametric horizontally-uniform micro-physical model. The parameters of this initialization are determined by a grid search of the cost function.
Furthermore, we added viewpoints corresponding to single-scattering angles, where polarization yields enhanced sensitivity to the droplet micro-physics (i.e., the cloudbow region). 
In addition, we introduce an optional adjustment, in which optimization of the liquid water content and effective radius are separated to alternating periods. 
The suggested initialization model and additional advances have been evaluated by retrieval of a set of large-eddy simulation clouds.

\end{abstract}


% \fi
\IEEEpeerreviewmaketitle

\section{Introduction}

\IEEEPARstart{T}{he} current state of the art remote sensing of clouds cannot produce sufficient information regarding the 3D nature of small warm clouds. This is both due to the resolution of most sensors currently in orbit, in which small clouds may be below pixel resolution, and even more importantly to the assumptions of the reconstruction algorithms. 

The plane-parallel assumption used in classic remote sensing effectively degenerates, to some extent, 3D light transfer to a 1D problem, as that there is effectively no horizontal light-transfer. The effect of this deficiency is particularly significant at the edges of the cloud. This leads to biased retrievals \cite{davis1997landsat,varnai2001statistical,zinner2006remote} and high uncertainties regarding the micro-physics and light-transfer of small clouds, regardless of the sensing resolution. 

To allow a better understanding of small-cloud micro-physics and light-transfer, a 3D remote sensing approach is essential. The development of such is currently receiving growing attention \cite{cornet2010three,marchand2004evaluation,marshak20053d,mayer2009radiative}.

Levis et al. \cite{levis2015airborne,Levis_2017_CVPR}, have developed a 3D scattering tomography algorithm, based on the Spherical Harmonic Discrete Ordinate Method (SHDOM) for radiative-transfer \cite{Evans1998}. Their method, pySHDOM \cite{levisCode}, retrieves cloud properties by fitting of multi-view light intensity images to a physics based forward-model.

This method has a scattering-based computational-tomography (CT) approach. It is a generalization of CT to recover scattering media such as clouds in a large scale. It is passive CT relying only on the Sun as an illumination source.  

The method was further developed using vSHDOM \cite{Doicu2013}, for vectorized radiative-transfer, allowing consideration of the polarization properties of light. In this approach, cloud properties are retrieved by fitting of the full Stokes vector images. Levis. et al. \cite{levis2020multi} have demonstrated by simulation the applicability of the method for multi-view cloud tomography.    

The polarization properties of light have advantages for retrieval of cloud-droplet size distribution \cite{Breon1998,Parol2004,Breon2005,Pust2007,Alexandrov2012,Shang2019,Sinclair2019}. However, these advantages are specifically limited to conditions where single-scattering may be assumed. For this reason, methods which use polarization for such retrievals have till recently been restricted to the top 50 m layer of clouds.

Recently, a new technique has been suggested, to allow evaluation of vertical profiles \cite{Alexandrov2020}. This technique uses the Research Scanning Polarimeter (RSP), not only for high angular resolution of the cloud top, but also to scan the sides of the cloud, allowing an approximation of a vertical profile of the cloud-droplet size distribution.

In this paper we complement the work of Levis et al. \cite{levis2020multi} by adjustments which are necessary for a realistic space-borne optical-imager. The simulated imager was modelled as having a polarized sensor the likes of Sony Polarsens.

We introduce a new method for initialization, based on a parameterized horizontally-uniform model. 

\section{Background}
\subsection{Parameter definition}
We define the cloud tomography problem as the retrieval of cloud-droplet micro-physics. For small clouds we assume droplets of spherical geometry\footnote{This is a common approximation for cloud-droplets, where Mie theory is used to describe light scattering.}, of different radius values $r$. We pursue the three-dimensional liquid water content (LWC), and droplet size distribution $n(r)$ per distance unit and volume unit to the extent of a chosen resolution. 

The LWC is defined, 
\begin{eqnarray}
   {\rm LWC} = \frac{4}{3} \pi  \rho_{\rm w}  \int_r \! \! r^3 n(r) {\rm d}r\;\;\;\;\;[\frac{\rm g}{{\rm m}^3}],
\label{eq:LWC}
\end{eqnarray}
where $\rho_{\rm w}=1\: \frac{\rm g}{{\rm cm}^3}$ is the water density.
Air molecular density is assumed to be known. Thus we do not try to retrieve air density.

To enable estimation of droplet size distribution, the distribution is compactly parameterized. Currently, we use a gamma distribution model to represent cloud-droplet size distribution ~\cite{hansen}, defined as follows:
\begin{eqnarray}
n(r) {=} N C r^{(v_{\rm e}^{-1} \! {-}3)} \exp [- r / (r_{\rm e}v_{\rm e})]\;\;\;\;\;[\frac{1}{\mu{\rm m}}\cdot\frac{1}{{\rm m}^3}],
\label{eq:gamma}
\end{eqnarray}
where $C{=} (r_{\rm e}v_{\rm e})^{(2{-}v_{\rm e}^{-1})} / \Gamma(v_{\rm e}^{-1} \! {-}2)$ is a normalization constant, $\Gamma$ is the Gamma function and $N$ is the droplet number concentration per volumetric unit
\begin{eqnarray}
 N = \int \limits_0^{\infty} n(r) {\rm d}r \;\;\;\;\;[\frac{1}{{\rm m}^3}].
\label{eq:N}
\end{eqnarray}
The effective radius, $r_{\rm e}\;\;[\mu {\rm m}]$, and unitless effective variance, $v_{\rm e}$ are defined by \cite{hansen}:
\begin{eqnarray}
 r_{\rm e} = \frac{\int (\pi r^2) r n(r) {\rm d}r}{\int ( \pi r^2) n(r) {\rm d}r}, \quad 
  v_{\rm e} = \frac{\int \left(r {-} r_{\rm e}\right)^2 (\pi r^2)n(r) {\rm d}r}{r_{\rm e}^2\int (\pi r^2) n(r) {\rm d}r}.
 \label{eq:reff_veff} 
\end{eqnarray} 

We currently focus on the retrieval of the LWC and $r_{\rm e}$ as in \cite{levis2020multi}.

\subsection{Retrieval quality measure}
The quality of each retrieval is quantitatively estimated by local mean error $\epsilon$ \cite{levis2020multi} defined:

\begin{equation}
     \epsilon_{\text{LWC}}=\frac{\|\hat{\text{LWC}}-\text{LWC}\|_1}{\|\text{LWC}\|_1}\:, \:
     \epsilon_{r_{\rm e}}=\frac{\|\hat{r_{\rm e}}-r_{\rm e}\|_1}{\|r_{\rm e}\|_1},
\label{eq:eps}
\end{equation}
where $\hat{\text{LWC}}$ and $\hat{r_e}$ are the estimated-value 3D fields, and ${\text{LWC}}$ and ${r_e}$ are the ground-truth values.

\section{Simulation overview}
\label{sec:overview}

A single simulation is constructed by the following stages:
\begin{enumerate}
    \item \textit{Measurments rendering}. Rendering of images by a forward model which runs a radiative-transfer simulation and then an image formation model. These are the simulation measurements (from here on, to avoid confusion, we define these images as the measurements). 
    This stage consists of the imager setup definition, and their optics (including noise). It also consists of definition of camera positions and views. With these settings, a set of images is rendered using the ground-truth micro-physical 3D fields. %The micro-physical 3D fields are simulated by Large-Eddy Simulations (LES).
    
    \item \textit{Initialization}. Definition of an initial state of the medium for optimization. 
    The initial state of the medium may be set in various ways. For instance, a medium grid at the beginning of  optimization may be entirely empty. However, the more similar the initial medium is to the ground-truth data, the better and faster the optimization will be. 
    
    \item \textit{Gradient descent based optimization}. In this stage, a set of images is again rendered by the forward model (stage 1). However, the model uses the current state (at a specific iteration) medium. These images are termed, the simulated images, which are used to fit the measurements. 
    
    A cost function is defined according to the gap between the simulated and measured images. 
\end{enumerate}

\section{Forward Model}
\label{sec:fmod}
This section describes the forward model which is used in the first and last stages of the simulation (see ~\ref{sec:overview}). It starts with the radiative-transfer simulation by pySHDOM as in \cite{Levis_2017_CVPR, levis2020multi, aides2020distributed}.
Then, the image formation model of a polarized camera is used to generate the multi-view images. Each pixel gathers the simulated Stokes radiance field from a set of origins and angles and finally transfer the Stokes radiance through polarizers of the camera. In addition, we implement a Cloudbow scan approach.

\subsection{Imager model}
\label{sec:Imager}

As in \cite{levis2020multi}, we use an imager model for both the forward and inverse models. We simulate a pinhole-camera model which is commonly used in computer-vision when the camera obeys {\em perspective projection}. The perspective projection is implemented in pySHDOM. Each pixel of the model's imager gathers radiance from the scene.  

Let us denote a spectral band by $\Lambda$. There, the wavelength $\lambda$ is between $[\lambda_1, \lambda_2]$. Let us denote a  spectral radiance at wavelength $\lambda$ by $I_{\lambda}$. It is calculated by pySHDOM and has units of $\big[ \frac{{\rm W}}{{\rm m}^2 \cdot {\rm sr} \cdot {\rm nm}} \big]$.

Assume a camera with lens of diameter $D$ at distance $f$ from the focal plane. The optics is perfect (no lens distortion), and the camera is focused on the object (i.e. clouds). Using pySHDOM, we simulate ${I}_{\lambda}$  that reaches the camera lens. Assume ${I}_{\lambda}$ is uniform within the area of a pixel footprint on the cloud. Geometrically, this is equivalent to saying that the irradiance on a detector pixel (area $p^2$) is uniform. For the rest of the text, we consider the pixels to be at optical axis of the camera\footnote{Note that if the pixel is not at the optical axis, there is a vignetting effect, which refers to radial fall-off of pixel intensity from the centre towards the edges of the image~\cite{goldman2010vignette}. For narrow field-of-view imagers, as we deal with, the vignetting is considered manageable.}.

To simulate a readout of the imager, we convert ${I}_{\lambda}$ within $\Lambda$ to photo-generated electrons in the sensor\footnote{For a sensor having a linear radiometric response, the conversion between electrons and the sensor readout value in gray-scales is by a fixed ratio. We do not deal with gray-scale values in this paper.}. A pixel on a sensor responds to the photons in a spectral band $\Lambda$. The pixel response depends on the sensor's quantum efficiency, $\mathrm{QE}_{\lambda} \: [\frac{\rm electrons}{\rm photons}]$, which is a measure of the probability for a photo-electron to be created per incident photon with wavelength $\lambda$.  Let $\tau_{\lambda}$ be the camera system efficiency due to optics losses and sensor reflection (not a part of $\mathrm{QE}$) at wavelength $\lambda$.  Light energy is converted to the expected number of photons at wavelengths
$\lambda$ by the factor $\frac{\lambda}{{\tt h} \: {\tt c}} \,[\frac{\rm photons}{\rm Joule}]$.

Let $\Delta t$ be the exposure time of the imager. 
The number of photo-electrons that are created in a pixel with a spectral band $\Lambda$ during exposure time $\Delta t$, is
\begin{align}
  N_{\lambda} &= p^2 \Delta t \int_{\Lambda} \tau_{\lambda} \mathrm{QE}_{\lambda} \, \frac{\lambda}{{\tt h} \, {\tt c}} \pi \big(\frac{D}{2f}\big)^2 {I}_{\lambda} {\rm d}\lambda \nonumber \\
  &= \Delta t \int_{\Lambda} \Gamma_{\lambda} {I}_{\lambda}  {\rm d}\lambda  \;\;\;\;\;[{\rm electrons}]. 
  \label{eq:N_Lambda}
\end{align}
Here we define
\begin{equation}
\label{eq:GAMMA}
    \Gamma_{\lambda} = 
    \pi \tau_{\lambda} \big(\frac{D}{2f}\big)^2 \mathrm{QE}_{\lambda} \, \frac{\lambda}{{\tt h} \, {\tt c}} p^2 \;\;\;\;\;
    \big[\frac{{\rm electrons} \cdot {\rm m}^2 \cdot {\rm sr}}{\rm Joule} \big],
\end{equation}
which encapsulates dependencies on the optics, pixel size, $\mathrm{QE}$ and the pixel's receptive solid-angle.\\

Ideally, to calculate the integral in Eq.~(\ref{eq:N_Lambda}), radiative-transfer would be run multiple times to calculate ${I}_{\lambda}$ in the spectral band in which $\Gamma_{\lambda}$ is significant (with certain wavelength resolution). There is common approximation that simplify these calculations \cite{Levis_2017_CVPR,aides2020distributed}. It is valid if wavelength dependencies within a spectral band are weak. This condition is met when narrow
bands are considered. In this approximation, the spectral radiance is considered fixed within $\Lambda$ and Eq.~(\ref{eq:N_Lambda}) becomes
\begin{align}
  N_{\Lambda} = \Delta t {I}_{\Lambda} \int_{\Lambda} \Gamma_{\lambda} L^{\mathrm{TOA}}_{\lambda} {\rm d}\lambda \;\;\;\;\; [{\rm electrons}].
  \label{eq:N_Lambda_final}
\end{align}
In Eq.~(\ref{eq:N_Lambda_final}), ${I}_{\Lambda}$\footnote{The radiative-transfer simulation is run on spectrally-averaged
 optical quantities \cite{Levis_2017_CVPR, aides2020distributed}.} is calculated with input solar irradiance of 1$\big[ \frac{{\rm W}}{{\rm m}^2 \cdot {\rm nm}} \big]$ and then scaled by the true solar irradiance $L^{\mathrm{TOA}}_{\lambda}$ [unitless] at the top of the atmosphere (TOA).

Now, we introduce a noise to $N_{\Lambda}$ to simulate a raw measurement of a camera. First we incorporate {\em photon shot noise}, which is Poisson-distributed around the $N_{\Lambda}$ expected value,
\begin{equation}
  N_{\Lambda}^{\rm measured} \sim  {\rm Poisson}\big\{ N_{\Lambda} \big\} \;\;\;\;\; [{\rm electrons}].
  \label{eq:N_Lambda_final_noise}
\end{equation} 
The maximum number of electrons which can be contained in a pixel is referred to as the {\em full well} capacity\footnote{Sensor suppliers specify it in units of [${\rm electrons}$].}. In our simulations, the exposure time $\Delta t$ is set to a level such that the sensor reaches 90\% of its full well. 

The sensor also introduces noise due to various causes, according to its specifications. The readout noise has a standard deviation of $\rho_{\rm read}$ [electrons]. The dark current shot noise at temperature $T$ is $D_T \: [\frac{\rm electrons}{\rm second}]$. The quantization noise is of standard deviation $\rho_{\rm digit}$ [electrons]. Overall, a pixel readout has a signal to noise ratio (SNR) of approximately 
\begin{equation}
  \text{SNR} =  \frac{N_{\Lambda}^{\rm measured}}{\sqrt{N_{\Lambda}^{\rm measured} + D_T \cdot \Delta t + \rho_{\rm read}^{2} + \rho_{\rm digit}^{2} }}.
  \label{eq:snr}
\end{equation} 
Noise specification in our simulation is based on Sony's IMX250MYR sensor \cite{sony_sensor}. The pixel size is $3.45 \times 3.45 \: {\rm \mu\text{m}}^2$, $\rho_{\rm read}$=2.31 [electrons], $D_T=3.51 \: [\frac{\text{electrons}}{\text{second}}]$ at 25 \textdegree{}${\rm C}$. A pixel reaches full-well with 10,500 electrons. We use 10 bit quantization. 

\subsection{Simulation of Stokes vector measurements}
\label{sec:Stokes_meas}
Our forward-model does not end with calculation of $N_{\Lambda}^{\rm measured}$. There is a need to adjust the forward-model output, to fit the inverse-model output, so that the cost function is based on the difference between the same quantities. These quantities are for instance, the three components of the Stokes vector, ${\bm S}_{\Lambda}=[I_{\Lambda},Q_{\Lambda},U_{\Lambda}]^{\rm T}$ that correspond to linear polarization (more differences defined in Sec.~\ref{sec:param_choise}). In both the forward and inverse models, each component should have units of  radiance.

The Sony IMX250MYR sensor \cite{sony_sensor} has four grid-wire polarizing filters which are formed on the chip in a block of four pixels. Each filter in a block has a different polarizing angle, either $90$\textdegree{}, $45$\textdegree{}, $135$\textdegree{}, or $0$\textdegree{}.

In pySHDOM, the Stokes vector is given in the meridian reference-frame. To simulate measurements of a polarized camera with four different polarization filters, we first need to convert the Stokes vector to a coordinate frame which is aligned with the camera reference-frame. Then to simulate the transmission of the converted Stokes vector through the polarizers.

We define the camera reference-frame so that it is aligned with the polarizer $0$\textdegree{} angle. The meridian reference-frame contains the view and zenith directions. Let $\hat{{\bm z}}$ denote the zenith direction vector at every point on Earth. In the Meridian reference-frame, the electric field components are defined by direction vectors \cite{levis2020multi}
\begin{eqnarray}
   {\bm b}  =  \frac{\hat{{\bm z}} \times {\bm \omega}}{ \| \hat{{\bm z}} \times {\bm \omega}\|}, \quad
   {\bm l} = {\bm \omega} \times {\bm b}.
\end{eqnarray}
where ${\bm \omega} = {\bm b}\times {\bm l}$ is a ray direction of a pixel and the axes ${\bm l}$, ${\bm b}$ are parallel and perpendicular to the the meridian plane (plane defined by ${\bm \omega}$ ~\cite{li2014method} and $\hat{{\bm z}}$), respectively. Let ${\bm S'}_{\Lambda}=[I'_{\Lambda},Q'_{\Lambda},U'_{\Lambda}]^{\rm T}$ be the Stokes vector in the camera reference-frame. The transformation between the meridian reference-frame to camera reference-frame is a multiplication of ${\bm S}_{\Lambda}$ by a Mueller rotation matrix 
\begin{eqnarray}
{\bf M} \left(\alpha \right) = 
    \begin{bmatrix} 
    1 & 0 & 0 & 0   \\
    0 & {\rm cos} \left(2\alpha \right) & {\rm -sin} \left(2\alpha \right) & 0   \\
    0 & {\rm sin} \left(2\alpha \right) & {\rm cos} \left(2\alpha \right) & 0   \\
    0 & 0  & 0 & 1  \\
    \end{bmatrix},
    \label{eq:phasemat}
\end{eqnarray}
where $\alpha$ is the rotation angle between ${\bm l}$ and the direction of linear polarizer with $0$\textdegree{} (calculated for each pixel individually), in the anti-clockwise direction when looking in the direction of propagation~\cite{li2014method}. Now
\begin{equation}
{\bm S'}_{\Lambda}={\bf M} \left(\alpha \right){\bm S}_{\Lambda},
\label{eq:Icam}
\end{equation}
and to go back to the meridian reference-frame,
\begin{equation}
{\bm S}_{\Lambda}={\bf M}^{-1} \left(\alpha \right){\bm S'}_{\Lambda}.
\label{eq:Icam_back}
\end{equation}

We simulate the transmission of ${\bm S'}_{\Lambda}$ through four pure linear polarizers with $90$\textdegree{}, $45$\textdegree{}, $135$\textdegree{} and $0$\textdegree{} angles. The transmitted radiances are
\begin{eqnarray}
    \begin{bmatrix} 
    I_{{\Lambda},90}   \\
    I_{{\Lambda},45}   \\
    I_{{\Lambda},135}   \\
    I_{{\Lambda},0}  \\
    \end{bmatrix} = 
    {\bf G} 
    \begin{bmatrix} 
    I'_{\Lambda}   \\
    Q'_{\Lambda}   \\
    U'_{\Lambda}   \\
    \end{bmatrix},
    \label{eq:g1}
\end{eqnarray}
where
\begin{eqnarray}
{\bf G}  = 
    \begin{bmatrix} 
    1 & {\rm cos} \left(2\psi_1 \right) & {\rm sin} \left(2\psi_1 \right)   \\
    1 & {\rm cos} \left(2\psi_2 \right) & {\rm sin} \left(2\psi_2 \right)   \\
    1 & {\rm cos} \left(2\psi_3 \right) & {\rm sin} \left(2\psi_3 \right)   \\
    1 & {\rm cos} \left(2\psi_4 \right) & {\rm sin} \left(2\psi_4 \right)  \\
    \end{bmatrix},
    \label{eq:g2}
\end{eqnarray}
and $\psi_1=90$, $\psi_2=45$, $\psi_3=135$, $\psi_4=0$.
On each radiance we apply the pipeline of Sec.~\ref{sec:Imager} to generate measurements $[N_{{\Lambda},90}^{\rm measured},N_{{\Lambda},45}^{\rm measured},N_{{\Lambda},135}^{\rm measured},N_{{\Lambda},0}^{\rm measured}]$. Then the inverse of Eq.~(\ref{eq:N_Lambda_final}) calculates $[I_{{\Lambda},90}^{\rm measured},I_{{\Lambda},45}^{\rm measured},I_{{\Lambda},135}^{\rm measured},I_{{\Lambda},0}^{\rm measured}]$. Finally the Stokes vector ${\bm S}_{\Lambda}^{\rm measured}=[I_{\Lambda}^{\rm measured},Q_{\Lambda}^{\rm measured},U_{\Lambda}^{\rm measured}]^{\rm T}$ is obtained by inverse of Eq.~(\ref{eq:g1}).
% \begin{eqnarray}
%     \begin{bmatrix} 
%     I_{\Lambda}^{\rm measured}   \\
%     Q_{\Lambda}^{\rm measured}   \\
%     U_{\Lambda}^{\rm measured}   \\
%     \end{bmatrix} = 
%     {\bf G}^{-1} 
%     \begin{bmatrix} 
%     I_{{\Lambda},90}^{\rm measured}   \\
%     I_{{\Lambda},45}^{\rm measured}   \\
%     I_{{\Lambda},135}^{\rm measured}   \\
%     I_{{\Lambda},0}^{\rm measured}  \\
%     \end{bmatrix}
%     \label{eq:g3}
% \end{eqnarray}
% , where $S_{\Lambda}^{\rm measured}=[I_{\Lambda}^{\rm measured},Q_{\Lambda}^{\rm measured},U_{\Lambda}^{\rm measured}]^{\rm T}$.
Than, convert ${\bm S}_{\Lambda}^{\rm measured}$ back to meridian reference-frame by Eq.~(\ref{eq:Icam_back}).\\
In the rest of the text, the band dependency is omitted for notation clarity.

\begin{figure}[t]
    \centering \includegraphics[width=0.9\linewidth]{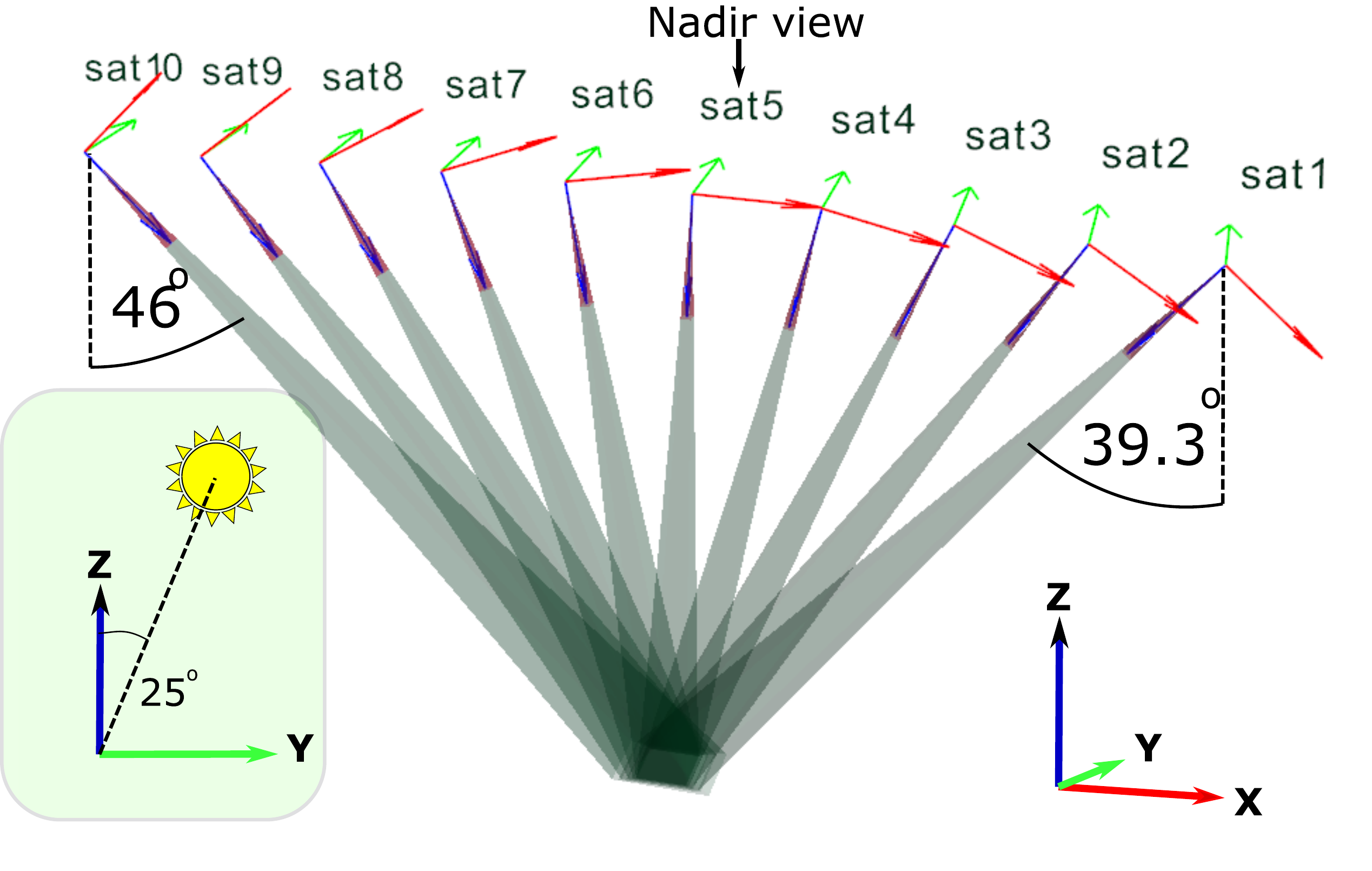}
    \caption {Imaging setup of 10 satellite at 500 km orbit. The East axis is labeled $y$, the North $X$, and the zenith $Z$. The satellites move in the positive direction of $x$ and $Y=0$. Each imager is aimed to the center of the cloud field. No satellite pointing error is introduced.  The distance between adjacent satellites is 100 km (on the orbit arc). The viewing angles are between $-46$\textdegree{} (sat 10) and $39.3$\textdegree{} (sat 1) relative to the zenith. The 10 satellites span a ~1000 km formation. Sat 5 views to the nadir. The solar zenith angle is
    $25$\textdegree{} and its azimuth angle is $90$\textdegree{} (the Sun irradiates from the East).}
    \label{fig:setup}
\end{figure}

Our imaging setup, described in Fig.~\ref{fig:setup}, requires imaging with high off-zenith viewing angles. It is a more challenging task than the setup in \cite{levis2020multi}, where the projection of each view was an orthographic projection. In contrast to orthographic projection, in perspective projection, images captured with high off-zenith view angles lose spatial resolution. The spatial resolution of the sensor influences the resolution of tomographic retrieval.

\subsection{Cloudbow scan}
\label{sec:Cloudbow}
\begin{figure*}  
\centering \includegraphics[width=16cm]{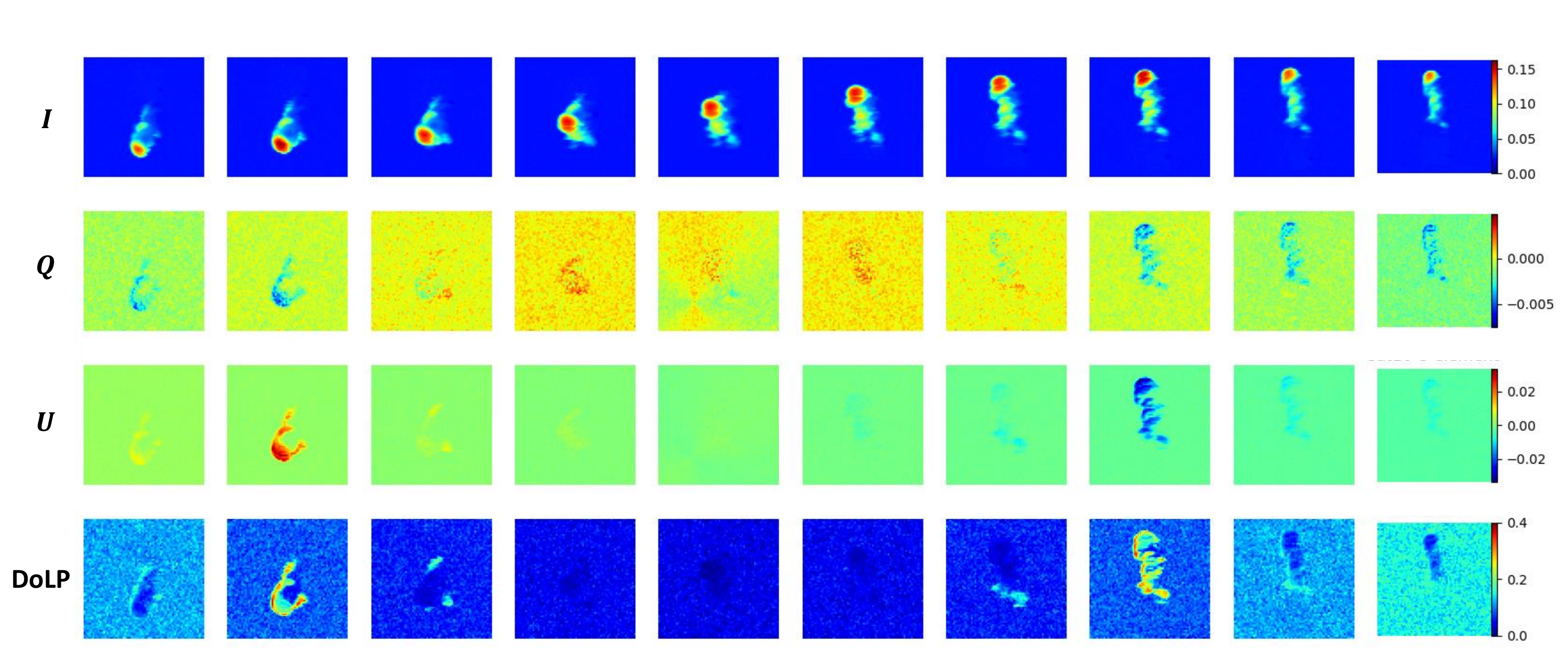}
\caption{The simulated (normalized) radiance I, Stokes elements Q, and U, and the DoLP, as viewed by the satellite setup described in Fig. \ref{fig:setup}. The leftmost images correspond to the view of satellite no. 1, the rightmost to that of satellite no. 10. The DoLP is visibly more noisy than the radiance. In addition, the high values of DoLP viewed by satellites no. 2 and no. 8, indicate the viewing angles of these satellites correspond to cloudbow scattering angles.}
\label{fig:stoke_images}
\end{figure*}

Mie scattering polarization is significant mostly in a specific range of scattering angles known as the cloudbow. The scattering angles of the cloudbow are bound approximately between $135$\textdegree, and $165$\textdegree (see Fig. \ref{fig:stoke_images}). The polarized intensity of scattering depends on droplet size distribution. In this domain, the polarization is highly sensitive to the effective radius of the droplets, as long as the effective variance is low \cite{Breon2005}. It is also sensitive to the wavelength of the scattered light. 

In order to better exploit the polarized information, a cloudbow-scanning principle is integrated into the process.
Additional images are obtained without changing the number of imagers, i.e. one or two satellites take more than one image of a single cloud field.

The idea is that one or two satellites capture additional images in viewing angles within the cloudbow, as illustrated in Fig. \ref{fig:stoke_images}. The angular resolution of sampling is between $1$\textdegree{} to $1.5$\textdegree{}, which is a realistic assumption for satellite attitude control. The algorithm we implemented finds these one or two satellites, per solar illumination direction.

 On average, we find the cloudbow scan decreases the mean errors $\epsilon_{\text{LWC}}$ and $\epsilon_{r_{\rm e}}$, by $2\%$ and $1\%$. We find this is a consistent improvement. However, further sensitivity study may improve the choice of scanned scattering angles and resolution.

\section{Initialization}

  \begin{figure*}[t]
\centering
\subfloat[]{\includegraphics[width=0.65\columnwidth]{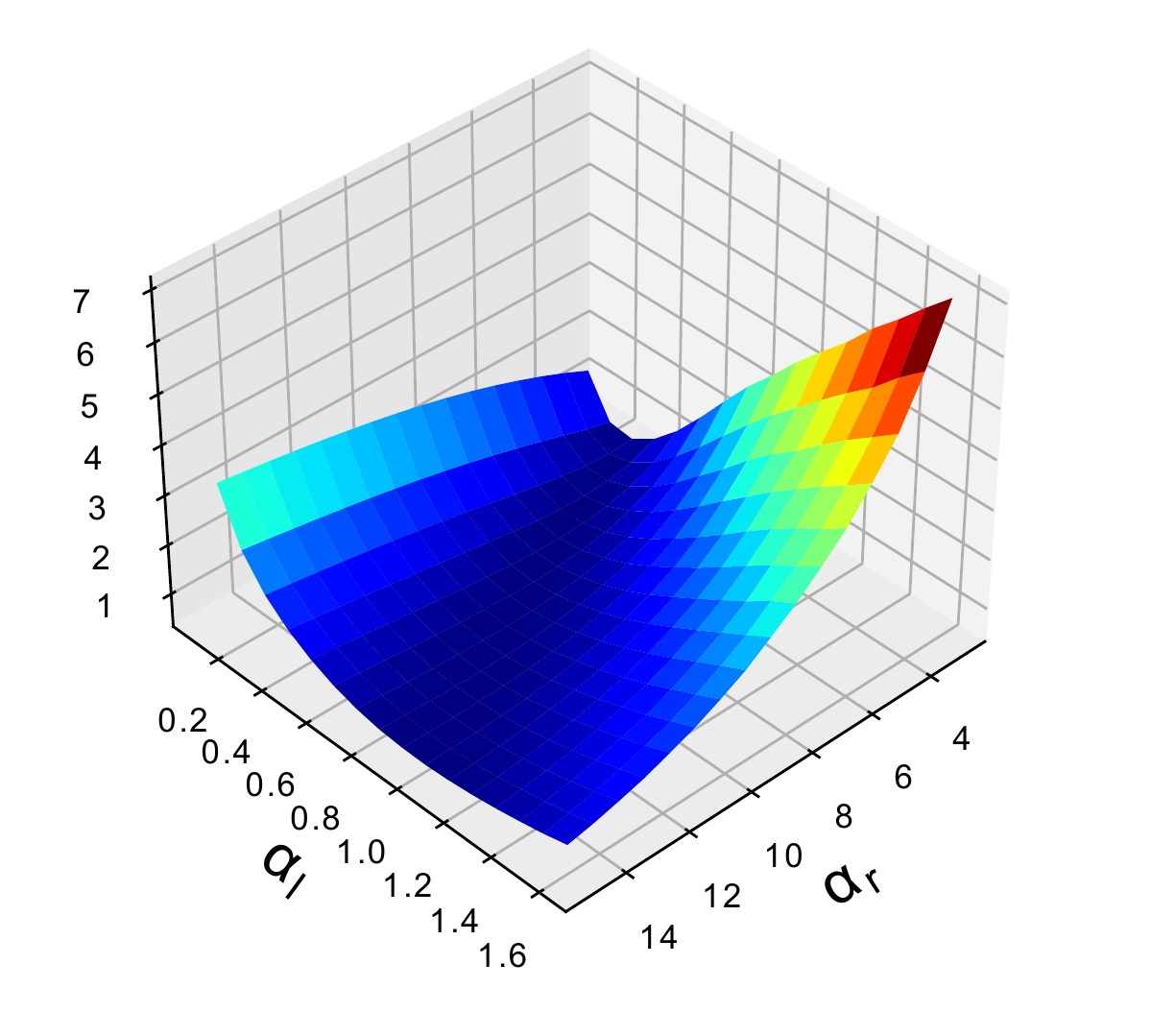}
\label{fig:I_cost}}
\subfloat[]{\includegraphics[width=0.65\columnwidth]{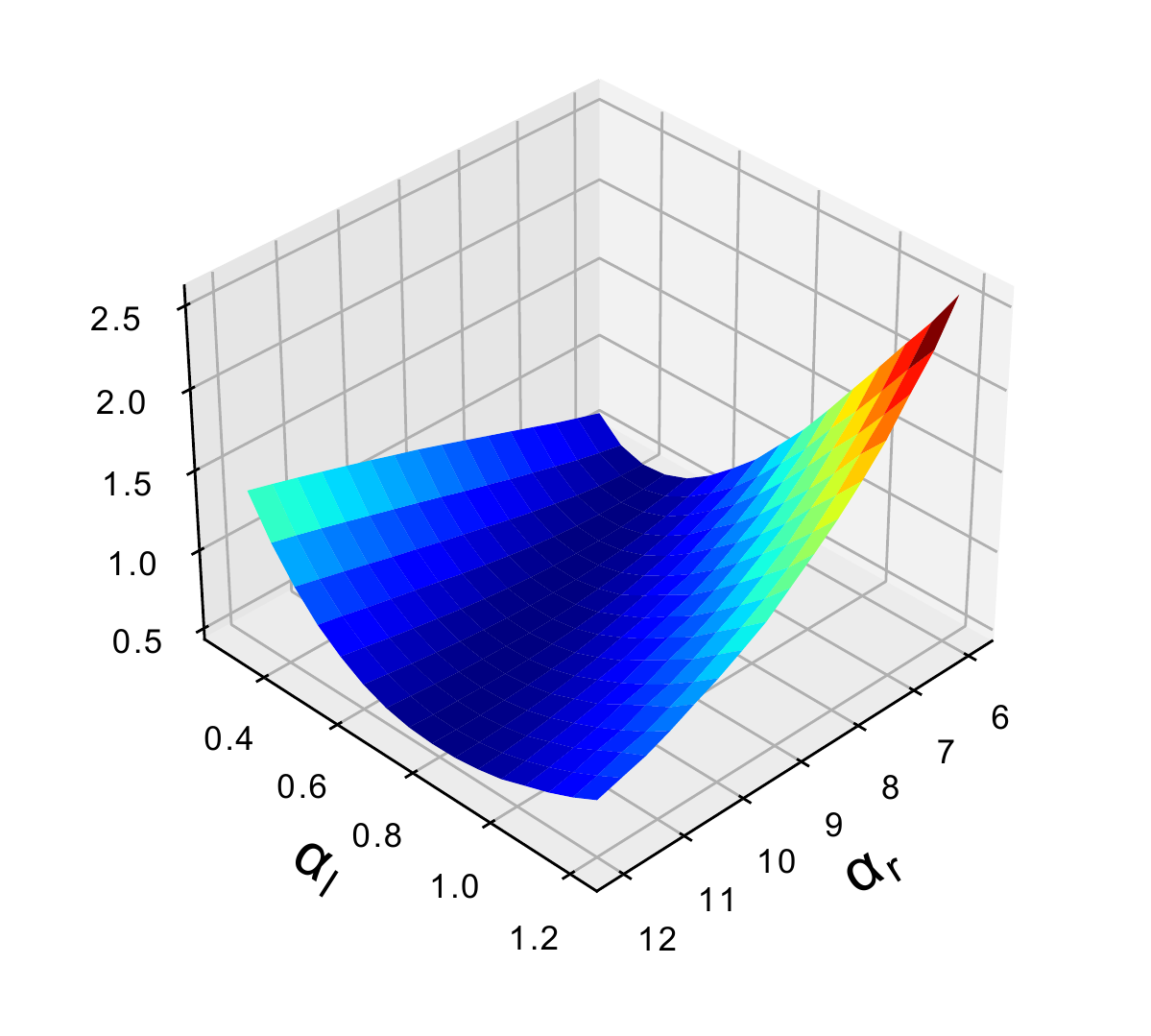}
\label{fig:stokes_cost}}
\subfloat[]{\includegraphics[width=0.65\columnwidth]{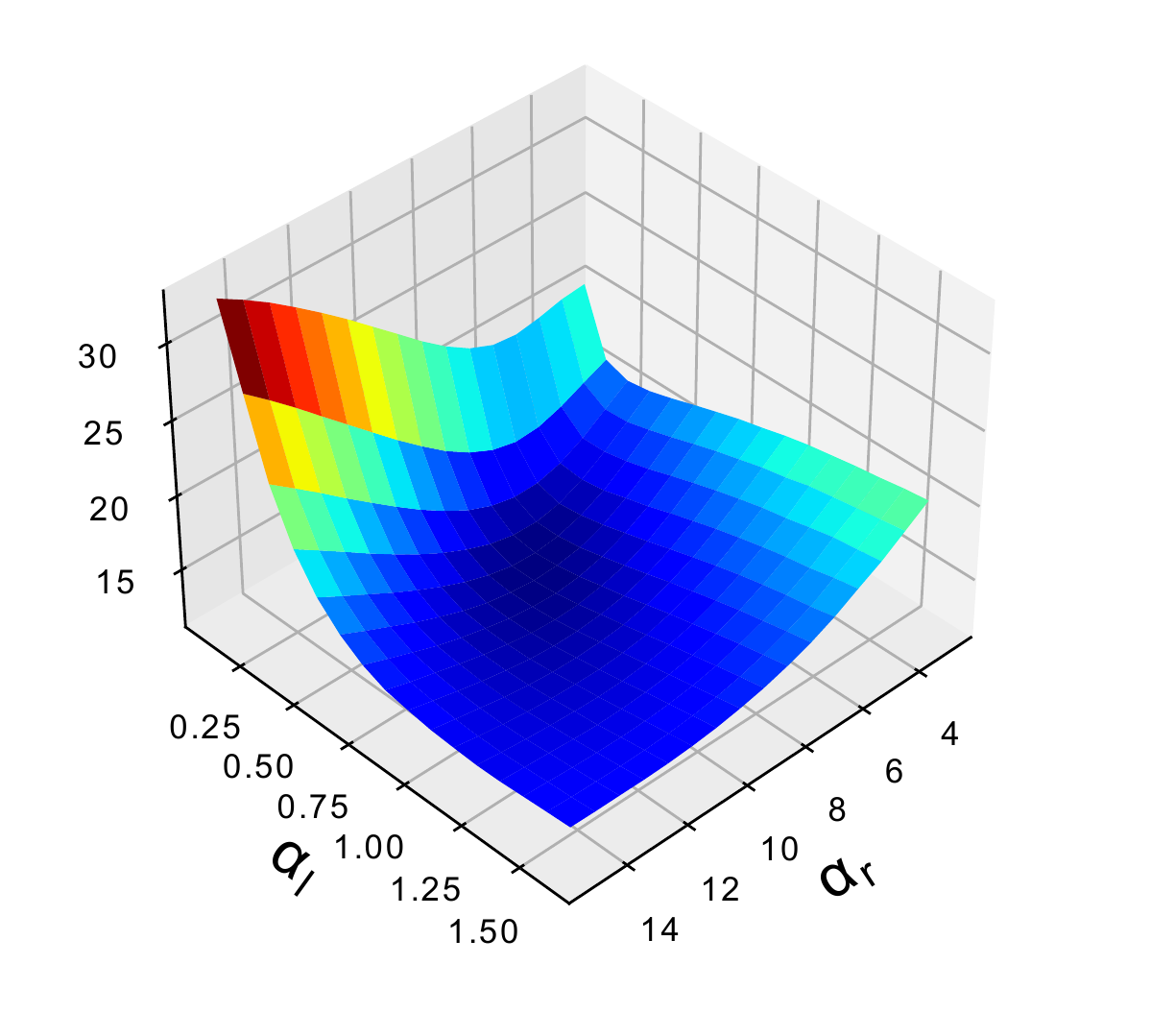}
\label{fig:dolp_cost}}
\caption{(a) $D_{\rm radiance}$, (b) $D_{\text{Stokes}}$, and (c) $D_{\text{DoLP}}$ as a function of initialization parameters. Note that the different scale of the cost values comes from different quantities differences among (a,b,c).}
\label{cost_surface}
\end{figure*}

This section describes the stage 2 of the simulation (see \ref{sec:overview}), definition of the initial medium for optimization. The cloud initialization has been found to be highly influential in gradient based CT optimization. The prior initialization method \cite{levis2020multi},  assumes a simple homogeneous model, using typical values of effective radius, $r_{\rm e}$, and LWC (in \cite{levis2020multi}, it is $r_{\rm e}=12 \: \mu{\text{m}}$, and $\text{LWC}=0.01 \: \frac{\text{g}}{\text{m}^3}$). We denote this method as $\text{H}_{\text{Typical}}$.

We now describe a horizontally-homogeneous parametric cloud model. The parameter values used for each initialization are found by optimization of the initial cost function, which would result from the parameter choice.

We emphasize the model is suggested strictly for initialization purposes. Once initialized, the optimization is no longer restricted to a horizontally-homogeneous assumption.

\subsection{Monotonous model}
The suggested cloud model assumes monotonic profiles of LWC and $r_e$ with altitude $Z\geq{Z_0}$ within a cloud, $Z_0$ being the cloud-base height \cite{loeub2020monotonicity}. The LWC is assumed to have a linear profile:
\begin{equation}
\text{LWC}\propto{(Z-Z_0)},    
\end{equation}
and $r_{\rm e}$ is assumed proportional to the cubic root of $Z$  \cite{loeub2020monotonicity}:
\begin{equation}
r_{\rm e}\propto{(Z-Z_0)^{\frac{1}{3}}}.    
\end{equation}

This assumption has been compared to Large-Eddy Simulations (LES) based on the Barbados Oceanographic and Meteorology Experiment (BOMEX), and is suitable for steady-state cumulus convection.
The profile equations set fixed minimum values for LWC ($\text{LWC}_{\rm min}$) and $r_{\rm e}$ ($r_{{\rm e}_{\rm min}}$). The profiles are defined as follows:
\begin{equation}
\text{LWC}=\alpha_{\rm l}(Z-Z_0)+\text{LWC}_{\rm min}  
\label{eq:lwc_par}
\end{equation}
\begin{equation}
r_{\rm e}=\alpha_{\rm r}(Z-Z_0)^{\frac{1}{3}}+ r_{{\rm e}_{\rm min}}   
\label{eq:reff_par}
\end{equation}
 Here $\alpha_{\rm l}$, and $\alpha_{\rm r}$ are slope parameters of the LWC and $r_{\rm e}$, respectively. Here $\text{LWC}_{\rm min}$ and $r_{{\rm e}_{\rm min}}$ are set to the values $0.0001 \: \frac{\text{g} d}{\text{m}^3}$, and $2.5 \: \mu{\text{m}}$, respectively. 
 
 The value is determined by the lower altitude of a 3D mask.\footnote{ Note that when using a mask determined by space-carving there is uncertainty in $Z_0$.}
 
 \subsection{Parameter choice}
\label{sec:param_choise}
Let $N_{\rm meas}$ be the length of the measurement vector, i.e. $N_{\rm meas} = N_{\rm view} N_{\rm pixels}$ (total number of viewpoints multiplied by a number of pixels of each view).
The slope parameters $\alpha_{\rm l}$, and $\alpha_{\rm r}$ used for a monotonous model are found by minimization of a cost function. We follow the notations of \cite{levis2020multi}.

For the specific purpose of optimizing the initialization, two cost function were considered for a polarized imaging. The first is defined by errors of all linear polarization components of the Stokes vector. For this purpose cost functions were defined for the components of the Stokes vector, ${\bm S}=[I,Q,U]^{\rm T}$, as the errors between the simulated components $S_i$, and the measured components $y_{S_i}$, at measurement index $k$ (pixel) and component index $i \in [1,3]$:
\begin{equation}
  D_{\rm radiance} = \frac{1}{2}\sum_{k=1}^{N_{\text{meas}}}(I[k]-y_I[k])^2,
\label{eq:cost_I}
\end{equation}
\begin{equation}
D_{Q} = \frac{1}{2}\sum_{k=1}^{N_{\text{meas}}}(Q[k]-y_Q[k])^2,
\label{eq:cost_Q}
\end{equation}
\begin{equation}  
D_{U} = \frac{1}{2}\sum_{k=1}^{N_{\text{meas}}}(U[k]-y_U[k])^2.
\label{eq:cost_U}
\end{equation}

The combined Stokes vector cost function, $D_{\rm Stokes}$ is defined as the sum of the separate component cost functions:  
\begin{equation}
D_{\rm Stokes} = D_{\rm radiance}+D_{Q}+D_{U}.
\label{eq:cost_S}
\end{equation}
 
 We define a second initialization method, $\text{M}_\text{Stokes}$, which uses the monotonous model, with parameters $\alpha_{\rm l}$, and $\alpha_{\rm r}$ set by a grid-scan search of $D_{\rm Stokes}$.
 
 The second considered cost is defined by the error of the degree of linear polarization (DoLP) where:
 \begin{equation}
     \text{DoLP}=\frac{\sqrt{Q^2+U^2}}{I}.
     \label{eq:dolp}
 \end{equation}
 We define the simulated DoLP at measurement index $k$:
  \begin{equation}
     \text{DoLP[k]}=\frac{\sqrt{Q[k]^2+U[k]^2}}{I[k]},
     \label{eq:dolp_measured}
 \end{equation}
 and the measured DoLP at measurement index $k$:
   \begin{equation}
     y_{\text{DoLP}}[k]=\frac{\sqrt{y_{Q}[k]^2+y_{U}[k]^2}}{y_{I}[k]}.
     \label{eq:dolp_simulated}
 \end{equation}
 Accordingly, the cost function is:
  \begin{equation}
   D_{\text{DoLP}} =  \frac{1}{2}\sum_{k=1}^{N_{\text{meas}}}(\text{DoLP}[k]-y_{\text{DoLP}}[k])^2.
   \label{eq:cost_dolp}
\end{equation}
 
 We define a third initialization method, $\text{M}_\text{DoLP}$, which uses the monotonous model, with parameters $\alpha_{\rm l}$, and $\alpha_{\rm r}$ set a by grid-scan search of $D_\text{DoLP}$.
 
\begin{figure*}[t]
\centering
\includegraphics[width=1.8\columnwidth]{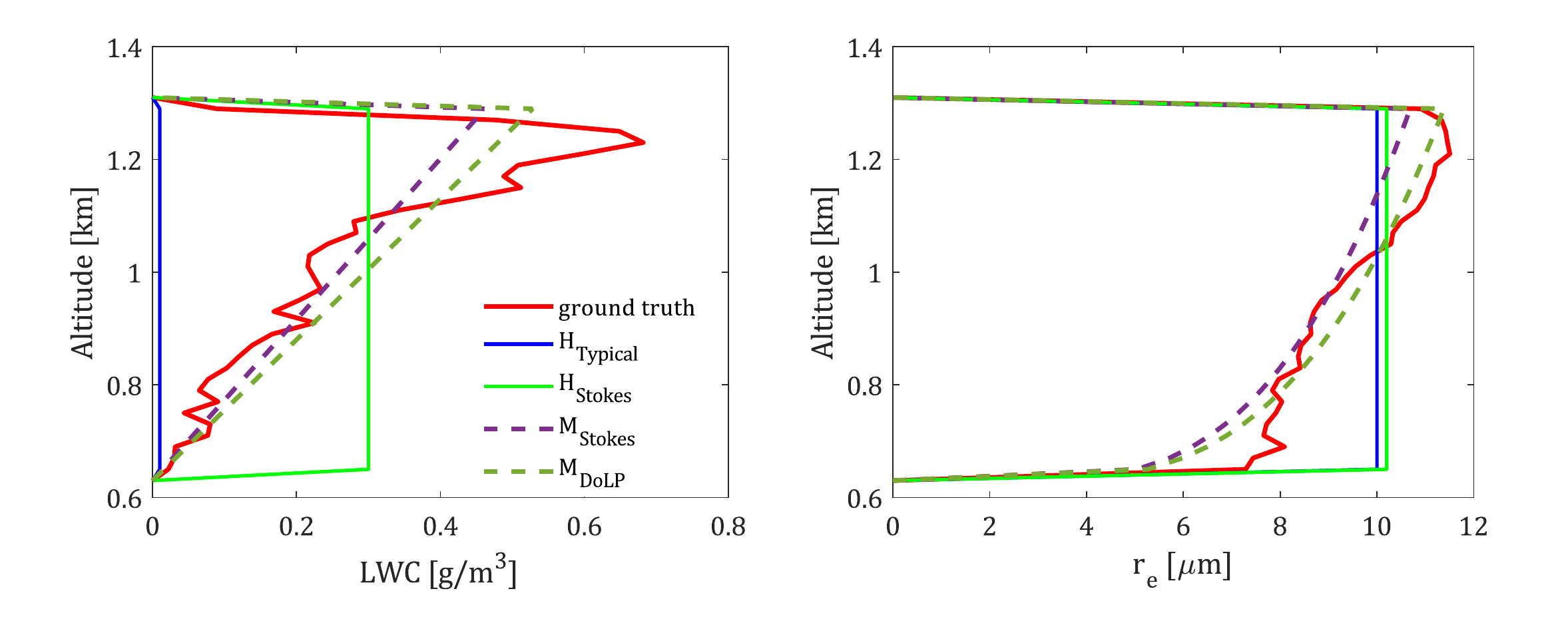}
\caption{Initial mean profiles of LWC and $r_{\rm e}$ for a sample cloud, as set by different initialization methods.}
\label{fig:Init_comparison}
\end{figure*}

 Examples of the cost function grid-planes are presented in Fig.~\ref{fig:Init_comparison}. The values of $D_{\rm radiance}$ in this example are a result of a simulated visible light camera. The values of $D_{Stokes}$ and $D_{\text{DoLP}}$ are a result of a simulated camera having a polarization sensor. 
 
 This example presents a potential advantage in using the DoLP for setting $\alpha_{\rm l}$ and $\alpha_{\rm r}$. The cost $D_{\text{DoLP}}$ shows a distinct minimum, while $D_{\text{Stokes}}$ results in a "valley" and is less explicit in defining the best parameters for initialization CT.

 For both $\text{M}_\text{Stokes}$ and $\text{M}_\text{DoLP}$ methods, the model parameters are set, by a grid scan. For $\alpha_{\text{l}}$, sixteen equally spaced values were considered between $0.1 \: \frac{\text{g}}{\text{m}^3\text{km}}$ and $1.6 \: \frac{\text{g}}{\text{m}^3\text{km}}$. For $\alpha_{r_{\rm e}}$, sixteen values were considered between $3 \: \frac{\mu\text{m}}{\text{km}}$ and $15 \: \frac{\mu\text{m}}{\text{km}}$. 
 
 The grid scan can also be used for a homogeneous model.  We define a fourth initialization method, $\text{H}_{\text{Stokes}}$, which uses the homogeneous model, with values LWC, and $r_{\rm e}$ set by grid-scan search of $D_{\rm Stokes}$.
 
 In this case the scan is over homogeneous values of the $\text{LWC}$ and ${r_{\rm e}}$. For comparison, we present retrieval results from this method as well. The grid search in this case is over the range of $\alpha_{\rm l}$ between $0.1 \: \frac{\text{g}}{\text{m}^3}$ and $1.6 \: \frac{\text{g}}{\text{m}^3}$ and $\alpha_{\rm r}$ values between $3 \: \mu\text{m}$ and $15 \: \mu\text{m}$.
 
 \begin{table}[h] 
  \caption{Summary of methods}
    \centering
    \begin{tabular}{l c}
    Method & Representation \\
    \hline\hline
    Homogeneous with typical values & $\text{H}_{\rm Typical}$ \\
    \hline
    Homogeneous with $D_{\rm Stokes}$  grid  search & $\text{H}_{\rm Stokes}$ \\
    \hline
     Monotonous with $D_{\rm Stokes}$  grid  search & $\text{M}_{\rm Stokes}$\\
    \hline
    Monotonous with $D_{\rm DoLP}$  grid  search & $\text{M}_{\rm DoLP}$ \\
    \hline
    \end{tabular}
    \label{tab:methods}
\end{table}

 An example of initialization mean profiles of $\text{LWC}$ and $r_{\rm e}$, as set by the different methods (summarized in Table \ref{tab:methods}),
 are presented in Fig. \ref{fig:Init_comparison}. 
 
 In this example, the model initialization of $\text{M}_{\text{Stokes}}$ and $\text{M}_\text{DoLP}$ yield different optimal initializations ($\alpha_{\rm l}$, $\alpha_{\rm r}$). This is not the case for all clouds. Anyway, conclusion regarding which initialization is better can be drawn only from resulting retrievals as described next.  

\section{Preconditioning and optimization in parts}
\label{sec:opt_in_parts}

This section relates to the third stage of the simulation (see \ref{sec:overview}.

The optimization attempts to solve a problem which consists of unknowns of two different orders of magnitude. In small clouds (up to a volume of $1\times1\times1\;\; \text{km}^3$), typical maximum values of LWC and $r_{\rm e}$  are $1\: \frac{\text{g}}{\text{m}^3}$, and $15 \: \mu\text{m}$, respectively. A gradient-descent based optimization method is highly affected by this relative scale. 

One way to overcome this effect is a preconditioning of the scales of the variables, which sets them at the same order of magnitude. Levis et al. \cite{levis2020multi} have shown the efficiency of this method, for preconditioning factors of $\Pi_{\text{LWC}} = 15$ and $\Pi_{r_{\rm e}} = 0.01$. However, we find that the preconditioning is case sensitive, and the ideal factors may vary greatly, depending on the characteristics of the cloud in question. For 3D variation in $r_{\rm e}$, we recommend preconditioning factors of $\Pi_{\text{LWC}} = 10$ and $\Pi_{r_{\rm e}} = 0.1$.

In addition, we propose an optional different approach. The gap in order of magnitude of each variable may be overcome by defining separate episodes of retrieval of the LWC and $r_{\rm e}$. The algorithm is as follows:
\begin{enumerate}
    \item Initialization of both LWC and $r_{\rm e}$.
    \item Gradient descent of LWC until convergence, while $r_{\rm e}$ is kept constant with values from initialization.
    \item Gradient descent of $r_{\rm e}$ until convergence, while LWC is kept constant with values from the previous estimation.
    \item Gradient descent of LWC until convergence, while $r_{\rm e}$ is kept constant with values from the previous estimation.
    \item Repeat stages 3 and 4 until the difference in the cost function between stages decreases below a set value. 
\end{enumerate}

\section{Simulations and Discussion}
The dataset used as ground-truth is a set of realistic 3D LWC and $r_{\rm e}$ arrays of six BOMEX-based LES-generated small clouds. The volumetric resolution of the ground-truth is $20\times{20}\times{20} \: {\text{m}^3}$. The effective variance $v_{\rm e}$ is assumed constant, $v_{\rm e}=0.1$. The characteristics of the cloud-set is summed in Table \ref{tab:clouds}. 

\begin{table} [b]
  \caption{Summary of ground-truth characteristics of cloud-set for simulations.}
    \centering
    \begin{tabular}{c c}
    \hline
    Maximum $r_{\rm e}$ [$\mu{\text{m}}$] & 8.8 - 15.4\\
    \hline
     Mean $r_{\rm e}$  [$\mu{\text{m}}$] & 6.2 - 10 \\
    \hline 
    Maximum cloud optical depth & 19.2 - 71.8 \\
    \hline
    Minimum cloud base altitude [\text{m}] & 550 \\
    \hline
    Maximum cloud top altitude [\text{m}] & 1710 \\
    \hline
    \end{tabular}
    \label{tab:clouds}
\end{table}

The configuration of satellites is constant in all demonstrations. A set of ten imagers in separate satellites is considered. The satellites are assumed to be in a string of pearls configuration, moving northward consecutively, as illustrated in Fig. \ref{fig:setup}. The altitude of the satellites is 500 km and the uniform distance between each consecutive pair is 100 km (on orbit arc). The viewing angles are between $-46$\textdegree{} and $39.3$\textdegree{} relative to the zenith. 

The coordinate system is local Cartesian coordinate system East, North, Up (ENU). In our convention, the East axis is labeled $Y$, the North $X$, and the zenith $Z$. The satellites move in the positive direction of $X$ and $Y=0$. Each imager is aimed to the center of the cloud field. No satellite pointing error is introduced. 

The angular position of the sun in the simulations is at 25\textdegree{} from the zenith, and at azimuth of 90\textdegree{}  (i.e., in the East).

The simulated imager has a narrow red waveband channel, between 620 nm and 670 nm. There, absorption by water droplets is negligible, and Rayleigh scattering is relatively low, reducing both airlight and sky polarization. The imager resolution is 20 m on the ground at nadir. 

The simulations include a cloudbow-scan. A single imager captures ten additional views within the scattering angles of [135\textdegree{}, 150\textdegree{}]. The retrieval assumes a ground-truth 3D mask, which is extracted from the 3D extinction field of the cloud-data.

\subsection{Initialization method}
For each initialization method, the $\epsilon$ error values of the cloud-set retrievals are averaged. The mean $\epsilon$ values of the different methods are compared.

\begin{figure}[t]
\centering
\includegraphics[width=1\columnwidth]{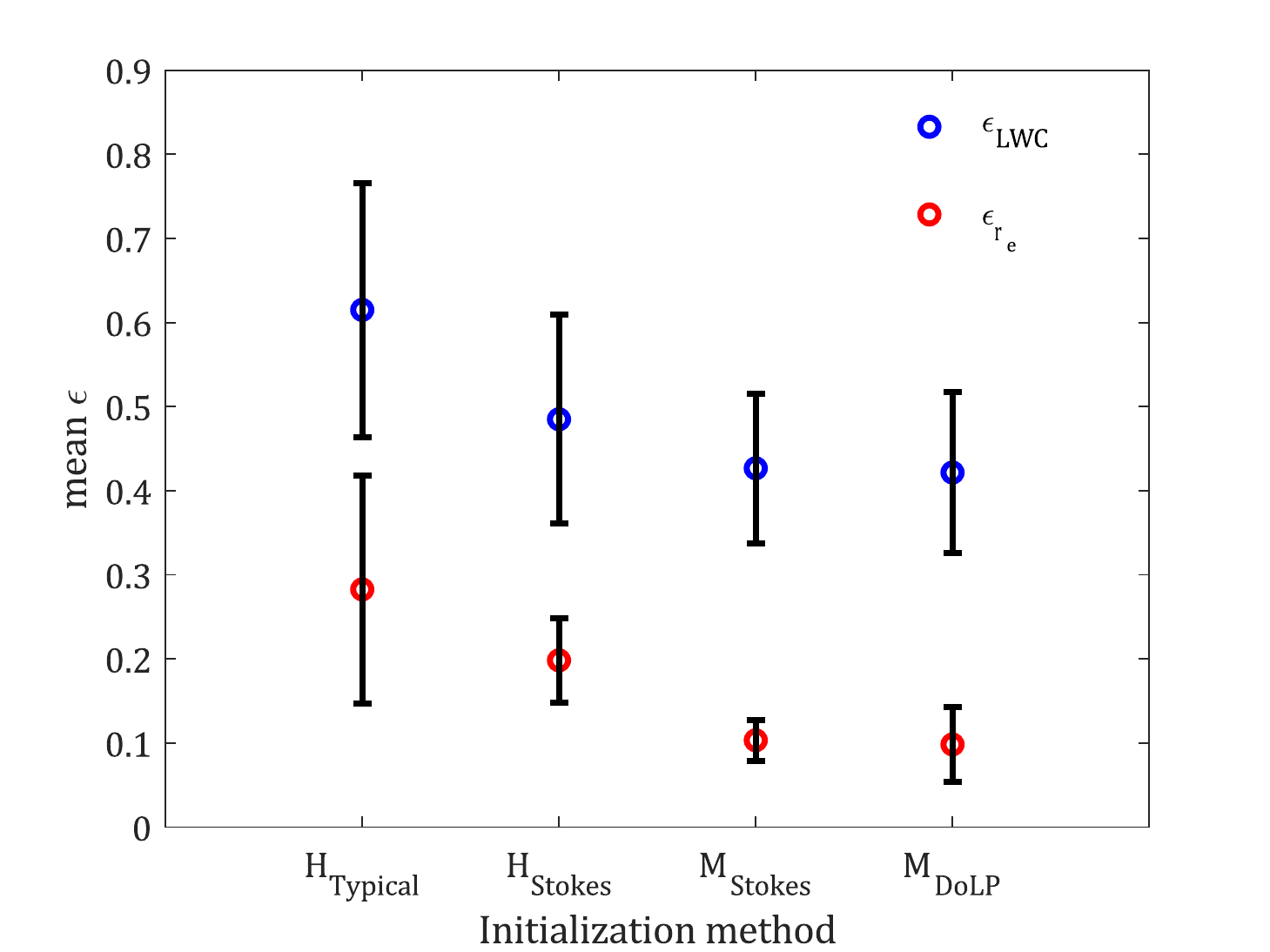}
\caption{Summary of mean error values for different initialization methods.}
\label{fig:summary}
\end{figure}

A comparison of the mean $\epsilon$ errors of the different methods (presented in Fig. \ref{fig:summary}), demonstrates the superiority of the $\text{M}_\text{Stokes}$ and $\text{M}_\text{DoLP}$ methods for initialization, which use the monotonous model.

 The $\text{H}_\text{Stokes}$ method is also an improvement with respect to $\text{H}_\text{Typical}$. Even so, the errors of both the LWC and $r_{\rm e}$ are even lower using the monotonous model methods.  

The difference in the mean of $\epsilon$ errors of $\text{M}_\text{Stokes}$ and $\text{M}_\text{DoLP}$ is very small, with a slight advantage to $\text{M}_\text{DoLP}$.

As mentioned in section \ref{sec:param_choise}, these two methods did not result in different initialization profiles for all cloud samples. For the few that yielded significantly different parameters we find an improvement in favor of the $\text{M}_\text{DoLP}$ method. For this reason, in spite of the seemingly small mean advantage, we recommend the use of $\text{M}_\text{DoLP}$ for the initialization method.
 
 \subsection{Optimization in parts}
 
 Assuming the $\text{M}_\text{DoLP}$ initialization method, we find that separating optimization periods for LWC and $r_{\text{e}}$ (see section \ref{sec:opt_in_parts}) yields an additional improvement to the retrieval. 
 
 The average errors decrease from $\epsilon_{\text{LWC}}=0.42$, $\epsilon_{r_{\text{e}}}=0.09$ to $\epsilon_{\text{LWC}}=0.39$, $\epsilon_{r_{\text{e}}}=0.08$.
  However, the separation does take a toll on the time of convergence, increasing it up to double.
 
 We find that for some cases the $r_{\rm e}$ optimization does but barely change the mean error. Even so, the LWC error may continue to decrease, in it's following optimization period. 
 
 We find that the initial mean profiles of $r_{\rm e}$, as set by $\text{M}_\text{DoLP}$, have very low initial errors.  Therefore when optimizing in parts, which begins with optimization of the LWC, conditions for LWC optimization are optimal. We notice two important implications in this respect. 
 
 First, beginning the optimization from $r_{\rm e}$ would not be as successful, as we have also observed (not presented here). Secondly, the 1D $r_{\rm e}$ model proves to be a good assumption, and the parameter grid search may be sufficient for $r_{\rm e}$ estimation. We may optionally reduce the optimization by gradient descent to estimation of LWC, while assuming a 1D $r_{\rm e}$ profile, retrieved by $\text{M}_\text{DoLP}$.

\section{Conclusion}

We have presented adjustments to the pySHDOM 3D micro-physical scattering tomography code. The major adjustments include implementation of a realistic polarized imager model, and a new initialization method. 

We have demonstrated the superiority of our proposed initialization method, under the conditions of the implemented imager model.

In addition, we have suggested two optional adjustments, cloudbow scanning, and optimization of the LWC and $r_{\rm e}$ in separate periods, which both have been demonstrated to reduce the mean error of the retrieval.

 %\begin{table*}[t]
 % \centering 
 % \caption{Summary of mean error values for different %initialization and run methods.}
 %  \begin{tabular}{c c c c c c} 
 %  \hline
 %  & Homogeneous & Homogeneous  & Model  & Model & Model 
 %  \\
 %  && + Stokes cost search  & + DoLP cost search & + DoLP cost %search & + DoLP cost search 
 %  \\
 %  && & &  no extra views & + Loops\\
 %  \hline
 %  $\epsilon{\text{LWC}}$ & $61.5 \pm 15.1$ & $48.5 \pm 12.4$ & %$42.1 \pm 9.5$ & $44.3 \pm 9.4$ & $39.6 \pm 9.8$ \\
 %  \hline
 %  $\epsilon{r_e}$ & $28.2 \pm 13.5$ & $19.8 \pm 5$ & $9.8 \pm %4.4$ & $11.1 \pm 2.8 $ & $8.3 \pm 1$ \\
 %  \hline
 %  \end{tabular}
 % \label{tab:summary}
%\end{table*}

% or


\appendices

\section*{Acknowledgment}
We are grateful to Ilan Koren, Orit Altaratz, Anthony Davis and Linda Forster for useful discussions and good advice. We thank Aviad
Levis and Jesse Loveridge for the pySHDOM code, for
useful discussions and being responsive to questions about it. We thank Johanan Erez, Ina Talmon and Daniel Yagodin for technical support.
Yoav Schechner is the Mark and Diane Seiden Chair in Science at the Technion. He is a Landau Fellow - supported
by the Taub Foundation. His work was conducted in the
Ollendorff Minerva Center. Minvera is funded through the
BMBF. This project has received funding from the European Union’s Horizon 2020 research and innovation programme under grant agreement No 810370-ERC-CloudCT

\ifCLASSOPTIONcaptionsoff
  \newpage
\fi



\bibliographystyle{IEEEtran}
\bibliography{ref}

\begin{thebibliography}{10}
\providecommand{\url}[1]{#1}
\csname url@samestyle\endcsname
\providecommand{\newblock}{\relax}
\providecommand{\bibinfo}[2]{#2}
\providecommand{\BIBentrySTDinterwordspacing}{\spaceskip=0pt\relax}
\providecommand{\BIBentryALTinterwordstretchfactor}{4}
\providecommand{\BIBentryALTinterwordspacing}{\spaceskip=\fontdimen2\font plus
\BIBentryALTinterwordstretchfactor\fontdimen3\font minus
  \fontdimen4\font\relax}
\providecommand{\BIBforeignlanguage}[2]{{%
\expandafter\ifx\csname l@#1\endcsname\relax
\typeout{** WARNING: IEEEtran.bst: No hyphenation pattern has been}%
\typeout{** loaded for the language `#1'. Using the pattern for}%
\typeout{** the default language instead.}%
\else
\language=\csname l@#1\endcsname
\fi
#2}}
\providecommand{\BIBdecl}{\relax}
\BIBdecl

\bibitem{davis1997landsat}
A.~Davis, A.~Marshak, R.~Cahalan, and W.~Wiscombe, ``The landsat scale break in
  stratocumulus as a three-dimensional radiative transfer effect: Implications
  for cloud remote sensing,'' \emph{Journal of the Atmospheric Sciences},
  vol.~54, no.~2, pp. 241--260, 1997.

\bibitem{varnai2001statistical}
T.~V{\'a}rnai and A.~Marshak, ``Statistical analysis of the uncertainties in
  cloud optical depth retrievals caused by three-dimensional radiative
  effects,'' \emph{Journal of the atmospheric sciences}, vol.~58, no.~12, pp.
  1540--1548, 2001.

\bibitem{zinner2006remote}
T.~Zinner and B.~Mayer, ``Remote sensing of stratocumulus clouds: Uncertainties
  and biases due to inhomogeneity,'' \emph{Journal of Geophysical Research:
  Atmospheres}, vol. 111, no. D14, 2006.

\bibitem{cornet2010three}
C.~Cornet, L.~C-Labonnote, and F.~Szczap, ``Three-dimensional polarized monte
  carlo atmospheric radiative transfer model (3dmcpol): 3d effects on polarized
  visible reflectances of a cirrus cloud,'' \emph{Journal of Quantitative
  Spectroscopy and Radiative Transfer}, vol. 111, no.~1, pp. 174--186, 2010.

\bibitem{marchand2004evaluation}
R.~Marchand and T.~Ackerman, ``Evaluation of radiometric measurements from the
  nasa multiangle imaging spectroradiometer (misr): Two-and three-dimensional
  radiative transfer modeling of an inhomogeneous stratocumulus cloud deck,''
  \emph{Journal of Geophysical Research: Atmospheres}, vol. 109, no. D18, 2004.

\bibitem{marshak20053d}
A.~Marshak and A.~Davis, \emph{3D radiative transfer in cloudy
  atmospheres}.\hskip 1em plus 0.5em minus 0.4em\relax Springer Science \&
  Business Media, 2005.

\bibitem{mayer2009radiative}
B.~Mayer, ``Radiative transfer in the cloudy atmosphere,'' in \emph{EPJ Web of
  Conferences}, vol.~1.\hskip 1em plus 0.5em minus 0.4em\relax EDP Sciences,
  2009, pp. 75--99.

\bibitem{levis2015airborne}
A.~Levis, Y.~Y. Schechner, A.~Aides, and A.~B. Davis, ``{Airborne
  three-dimensional cloud tomography},'' in \emph{Proceedings of the IEEE
  International Conference on Computer Vision}, 2015, pp. 3379--3387.

\bibitem{Levis_2017_CVPR}
A.~Levis, Y.~Y. Schechner, and A.~B. Davis, ``Multiple-scattering microphysics
  tomography,'' in \emph{Proceedings of the IEEE Conference on Computer Vision
  and Pattern Recognition (CVPR)}, July 2017.

\bibitem{Evans1998}
K.~F. Evans, ``{The spherical harmonics discrete ordinate method for
  three-dimensional atmospheric radiative transfer},'' \emph{Journal of the
  Atmospheric Sciences}, vol.~55, no.~3, pp. 429--446, 1998.

\bibitem{levisCode}
A.~Levis and A.~Aides, ``pyshdom,''
  \url{https://github.com/aviadlevis/pyshdom}, 2019.

\bibitem{Doicu2013}
\BIBentryALTinterwordspacing
A.~Doicu, D.~Efremenko, and T.~Trautmann, ``{A multi-dimensional vector
  spherical harmonics discrete ordinate method for atmospheric radiative
  transfer},'' \emph{Journal of Quantitative Spectroscopy and Radiative
  Transfer}, vol. 118, pp. 121--131, 2013. [Online]. Available:
  \url{http://dx.doi.org/10.1016/j.jqsrt.2012.12.009}
\BIBentrySTDinterwordspacing

\bibitem{levis2020multi}
A.~Levis, Y.~Y. Schechner, A.~B. Davis, and J.~Loveridge, ``Multi-view
  polarimetric scattering cloud tomography and retrieval of droplet size,''
  \emph{Remote Sensing}, vol.~12, no.~17, p. 2831, 2020.

\bibitem{Breon1998}
F.-M. Breon and P.~Goloub, ``{Cloud droplet effective radius from spaceborne
  polarization measurements},'' \emph{Geophysical Research Letters}, vol.~25,
  no.~11, pp. 1879--1882, 1998.

\bibitem{Parol2004}
F.~Parol, J.~C. Buriez, C.~Vanbauce, J.~Riedi, L.~C.-Labonnote,
  M.~Doutriaux-Boucher, M.~Vesperini, G.~S{\`{e}}ze, P.~Couvert, M.~Viollier,
  and F.~M. Br{\'{e}}on, ``{Review of capabilities of multi-angle and
  polarization cloud measurements from POLDER},'' \emph{Advances in Space
  Research}, vol.~33, no.~7, pp. 1080--1088, 2004.

\bibitem{Breon2005}
F.~M. Br{\'{e}}on and M.~Doutriaux-Boucher, ``{A comparison of cloud droplet
  radii measured from space},'' \emph{IEEE Transactions on Geoscience and
  Remote Sensing}, vol.~43, no.~8, pp. 1796--1805, 2005.

\bibitem{Pust2007}
N.~J. Pust, J.~A. Shaw, A.~Sinyuk, O.~Dubovik, B.~Holben, T.~F. Eck, F.~M.
  Breon, J.~Martonchik, R.~Kahn, D.~J. Diner, E.~F. Vermote, J.~C. Roger,
  T.~Lapyonok, I.~Slutsker, J.~C. Beckert, T.~H. Reilly, C.~J. Bruegge, J.~E.
  Conel, R.~A. Kahn, J.~V. Martonchik, T.~P. Ackerman, R.~Davies, S.~A. {W
  Gerstl}, H.~R. Gordon, J.~P. Muller, R.~B. Myneni, P.~J. Sellers, B.~Pinty,
  M.~M. Verstraete, B.~N. Holben, D.~Tanre, J.~P. Buis, A.~Setzer, E.~Vermote,
  J.~A. Reagan, Y.~J. Kaufman, T.~Nakajima, F.~Lavenu, I.~Jankowiak,
  A.~Smirnov, G.~P. Anderson, P.~K. Acharya, L.~S. Bernstein, L.~Muratov,
  J.~Lee, M.~Fox, S.~M. Adler-Golden, J.~H. Chetwynd, M.~L. Hoke, R.~B.
  Lockwood, J.~A. Gardner, T.~W. Cooley, C.~C. Borel, and P.~E. Lewis,
  ``{Wavelength dependence of the degree of polarization in cloud-free skies:
  simulations of real environments " Multi-angle Imaging SpectroRadiometer
  (MISR) instrument description and experiment overview},'' \emph{IEEE Trans.
  Geosci. Rem. Sens. OPTICS EXPRESS}, vol. 107, no.~14, pp. 479--507, 2007.

\bibitem{Alexandrov2012}
\BIBentryALTinterwordspacing
M.~D. Alexandrov, B.~Cairns, C.~Emde, A.~S. Ackerman, and B.~van Diedenhoven,
  ``{Accuracy assessments of cloud droplet size retrievals from polarized
  reflectance measurements by the research scanning polarimeter},''
  \emph{Remote Sensing of Environment}, vol. 125, pp. 92--111, 2012. [Online].
  Available: \url{http://dx.doi.org/10.1016/j.rse.2012.07.012}
\BIBentrySTDinterwordspacing

\bibitem{Shang2019}
\BIBentryALTinterwordspacing
H.~Shang, H.~Letu, F.~M. Br{\'{e}}on, J.~Riedi, R.~Ma, Z.~Wang, T.~Y. Nakajima,
  Z.~Wang, and L.~Chen, ``{An improved algorithm of cloud droplet size
  distribution from POLDER polarized measurements},'' \emph{Remote Sensing of
  Environment}, vol. 228, no. April, pp. 61--74, 2019. [Online]. Available:
  \url{https://doi.org/10.1016/j.rse.2019.04.013}
\BIBentrySTDinterwordspacing

\bibitem{Sinclair2019}
\BIBentryALTinterwordspacing
K.~Sinclair, B.~van Diedenhoven, B.~Cairns, M.~Alexandrov, R.~Moore,
  E.~Crosbie, and L.~Ziemba, ``{Polarimetric retrievals of cloud droplet number
  concentrations},'' \emph{Remote Sensing of Environment}, vol. 228, no. August
  2018, pp. 227--240, 2019. [Online]. Available:
  \url{https://doi.org/10.1016/j.rse.2019.04.008}
\BIBentrySTDinterwordspacing

\bibitem{Alexandrov2020}
\BIBentryALTinterwordspacing
M.~D. Alexandrov, D.~J. Miller, C.~Rajapakshe, A.~Fridlind, B.~van Diedenhoven,
  B.~Cairns, A.~S. Ackerman, and Z.~Zhang, ``{Vertical profiles of droplet size
  distributions derived from cloud-side observations by the research scanning
  polarimeter: Tests on simulated data},'' \emph{Atmospheric Research}, vol.
  239, no. February, p. 104924, 2020. [Online]. Available:
  \url{https://doi.org/10.1016/j.atmosres.2020.104924}
\BIBentrySTDinterwordspacing

\bibitem{hansen}
J.~E. Hansen and L.~D. Travis, ``Light scattering in planetary atmospheres,''
  \emph{Space science reviews}, vol.~16, no.~4, pp. 527--610, 1974.

\bibitem{aides2020distributed}
A.~Aides, A.~Levis, V.~Holodovsky, Y.~Y. Schechner, D.~Althausen, and
  A.~Vainiger, ``Distributed sky imaging radiometry and tomography,'' in
  \emph{2020 IEEE International Conference on Computational Photography
  (ICCP)}.\hskip 1em plus 0.5em minus 0.4em\relax IEEE, 2020, pp. 1--12.

\bibitem{goldman2010vignette}
D.~B. Goldman, ``Vignette and exposure calibration and compensation,''
  \emph{IEEE transactions on pattern analysis and machine intelligence},
  vol.~32, no.~12, pp. 2276--2288, 2010.

\bibitem{sony_sensor}
``Polarization image sensor with four-directional on-chip polarizer and global
  shutter function,''
  \url{https://www.sony-semicon.co.jp/e/products/IS/industry/product/polarization.html}.

\bibitem{li2014method}
L.~Li, Z.~Li, K.~Li, L.~Blarel, and M.~Wendisch, ``A method to calculate stokes
  parameters and angle of polarization of skylight from polarized cimel sun/sky
  radiometers,'' \emph{Journal of Quantitative Spectroscopy and Radiative
  Transfer}, vol. 149, pp. 334--346, 2014.

\bibitem{loeub2020monotonicity}
T.~Loeub, A.~Levis, V.~Holodovsky, and Y.~Y. Schechner, ``Monotonicity prior
  for cloud tomography,'' in \emph{Proceedings of the European Conference on
  Computer Vision (ECCV), Glasgow, Scotlang}, 2020, pp. 24--29.

\end{thebibliography}


% \end{IEEEbiography}


%\newpage



%\vfill


\end{document}